\documentstyle[12pt]{article}
\newfont {\xx} {cmti10}

\begin{document}
\def\smp{Standard Model. }
\def\sm{Standard Model }
\def\be{\begin{equation}}
\def\l{\label}
\def\r{\ref}
\def\ee{\end{equation}}
\def\bea{\begin{eqnarray}}
\def\eea{\end{eqnarray}}
\def\nn{\nonumber \\}
\def \R{(\frac{ \alpha_i(0)}{ \alpha_i(t)})}
\def \RI{(\frac{\alpha_i(t)}{\alpha_i(0)})}
\def \R3{(\frac{\alpha_3(0)}{\alpha_3(t)})}
\def \RI3{(\frac{\alpha_3(t)}{\alpha_3(0)})}
\def \EP{($\frac{\alpha(t)}{\alpha(0)})^B$ }
\def \EPP{($\frac{\alpha(M_X)}{\alpha(M_C)})^B$ }
\def \FPR{($\frac{\tilde\alpha(0)}{Y_t(0)})/
(\frac{\tilde\alpha}{Y_t})^*$}
\def \sc2e{\sum_iC_2(R_i)}
\def \sc2{$\sum_iC_2(R_i)$}
\def\c2e{C_2(R_i)}
\def\c2{$C_2(R_i)$}
\def\ta3{\tilde\alpha_3}
\def\fr{\frac{16}{3}}
\def\at{\frac{A(t)}{M(t)}}
\def\xt{\frac{X(t)}{M(t)^2}}
\def\mqt{\frac{m_q^2(t)}{M(t)^2}}
\def\mht{\frac{\mu_2^2}{M(t)^2}}
\def\mtt{\frac{m_t^2(t)}{M(t)^2}}
\def\xtmu{\frac{X(t)}{\mu^2(t)}}
\def\mqtmu{\frac{m_q^2(t)}{\mu^2(t)}}
\def\mhtmu{\frac{\mu_2^2(t)}{\mu^2(t)}}
\def\b0{\frac{B}{M}}
\def\aa{\frac{A}{M}}
\def\a2m{\frac{A^2}{M^2}}
\def\mq{\frac{m_q^2}{M^2}}
\def\mt{\frac{m_t^2}{M^2}}
\def\mh{\frac{\mu_2^2 - \mu^2}{M^2}}
\def\b0mu{\frac{B}{\mu}}
\def\amu{\frac{A}{\mu}}
\def\a2mu{\frac{A^2}{\mu^2}}
\def\ammu{\frac{A M}{\mu^2}}
\def\mqmu{\frac{m_q^2}{\mu^2}}
\def\mtmu{\frac{m_t^2}{\mu^2}}
\def\mhmu{\frac{\mu_2^2}{\mu^2}}
\def\y{(Y_0-Y^*)}
\def\mmu{\frac{M^2}{\mu^2}}
\title{Infra-red fixed point structure of soft  supersymmetry
breaking
mass terms}
\author{Marco Lanzagorta$^a$, Graham G. Ross$^b$\thanks{SERC
Senior
Fellow, on leave from $^a$},\\
$^a $Department of Physics,
Theoretical Physics,\\
University of Oxford,
1 Keble Road,
Oxford OX1 3NP\\  $^b$ Theory Division, CERN, CH-1211,\\ Geneva,
Switzerland}

\date{}
\maketitle
\vspace{-10cm}
\hspace{13cm}
CERN-TH-161
\vspace{10cm}

\abstract{{\small{We show that the soft SUSY breaking mass terms may
have
infra-red stable fixed points at which they are related to the
gaugino
masses and argue that in a generic unification these masses should
lie
 close to their fixed points. We consider the implications for the
family dependence of squark and slepton masses and the related
flavour
changing neutral currents and determine conditions under which models
with flavour changing couplings and masses at a high scale may lead
to
a family independent effective theory at low scales. The analysis is
illustrated for a variety of models for which we compute both the
fixed
point structure and determine the rate of approach to the fixed
point.}}

\bigskip
\noindent{\bf Introduction}
\bigskip

Our knowledge of the fundamental laws of nature are entirely
consistent
with the idea that what we are observing is an effective low energy
theory following from some underlying stage of unification at a very
high scale $M_X$. Provided we add a stage of low energy supersymmetry
all the general features of the \sm including spontaneous symmetry
breakdown can be understood. The light states of the theory are those
that are protected from the high scale by symmetry; fermions are
chiral
and vectors are associated with a local gauge symmetry. The
supersymmetry allows us to
have light scalars provided they are associated by supersymmetry to
chiral fermions. Given this very attractive picture it is natural to
ask whether the low energy parameters of the theory giving the
couplings masses and mixing angles can similarly be understood on the
basis of the low-energy properties of the theory. The reason this may
be possible is that, although the fundamental couplings should be
determined by the underlying theory, their measured values correspond
to their low-energy values. The renormalisation group equations which
are needed to determine the low-energy values arecompletely
determined in terms of the low-energy structure of the theory. Thus
the
initial values of the masses and  couplings determined by the
underlying unified(?) theory  serve only to give the boundary
conditions for the parameters of the effective low-energy theory.
This
raises the possibility that the low energy parameters may be largely
determined by the dynamics of the low-energy theory itself through
the
infra-red fixed point structure of its renormalisation group
equations,
the value of a parameter (or ratio of parameters) at the fixed point
being insensitive to the initial value.

In the context of the \sm the only ratio of parameters that
 may lie close to an (approximate) infra red stable fixed point
(IRSFP)
 is the ratio of the top Yukawa coupling to the QCD gauge
 coupling\cite{pr}\footnote{Only in the absence of the $SU(2)\otimes
U(1)$
factors
is there a true fixed point.}. In the minimal supersymmetric
extension
of
 the model (MSSM) the ratio of both the top and bottom Yukawas may
lie
close to their fixed point values. In a previous paper \cite{lr} we
discussed
 this possibility in some detail and computed how close to the fixed
point the ratios are likely to be driven in going from the
unification
scale to the scale of electroweak breaking. We also raised the
question
whether the remaining couplings of the theory might be determined by
the fixed points of the underlying theory lying beyond the \sm
because
we found that the rate of approach to the fixed point could be very
rapid. While this does not realise the original aim of interpreting
couplings as due to the dynamics at {\it low} scales it does raise
the
possibility that the couplings are determined by the dynamics of the
effective theory below the Planck or string scale. The beauty of the
idea is that the structure of the RG equations are determined by the
multiplet structure and symmetries of the theory, features that it
may
be possible to determine from the structure of the theory at the
electroweak scale. For example the pattern of fermion masses and
mixings strongly suggests the existence of a (spontaneously broken)
family symmetry above the gauge unification scale \cite{fn}.
Identification of this symmetry is sufficient to determine the RG
equations and thus, possibly, to completely determine the mass
matrices
via the IRSFP structure.  In this paper we extend the discussion of
the
fixed point structure to include the  implications for the soft SUSY
breaking terms. This allows us to address the problem of flavour
violation associated with any attempt to enlarge the symmetries of
the
\sm to include a family symmetry. For closely related work on the
infrared structure of such soft terms see refs. \cite{st2,j1}.

  The general form of
the effective scalar potential with soft SUSY breaking is:

\bea
V_{eff} &=& \sum_{i} \left| \frac{\partial \tilde W}{\partial Z_i}
\right|^2 +\sum_{i} |m_i|^2  |Z_i|^2 + \sum_{j}(( A _ j\tilde W_{3j}+
h.c. ) \nn
        & & +  ( B_j \tilde W_{2j} + h.c. ) + ( C_j \tilde W_{j} +
h.c.
))
            + \hbox{gauge terms}
\l{eq:1}
\eea
where $ \tilde W$ is an effective low-energy superpotential and
$ \tilde W_{3j}$, $\tilde W_{2j}$ and $ \tilde W_{1j} $ are the terms
making up its trilinear,
bilinear and linear parts. The parameters $m_i,\;A_i,\;B_i$ and $C_i$
are the soft SUSY breaking terms.

\bigskip
\noindent{\bf IRSFP structure.}
\bigskip

To discuss the implications for these terms following from the IRSFP
structure we start with the RG equation for the case there is a
single
gauge coupling corresponding to a single gauge group factor (as is
the
case for $SU(5),\;SO(10)$ or as an approximation in the case of the
MSSM) or a product of identical gauge groups with a permutation
symmetry (e.g. $SU(3)^3$). To illustrate the general features we
first
consider the simple case where there is only a single generation and
a
single Yukawa coupling $hQ_LQ_R H$ (such as is the case if  one
coupling dominates). We will consider more general couplings shortly.
The RG equations in this case are
\be
\frac{d \tilde\alpha}{d t}=-b \tilde\alpha^2
\label{eq:2}
\ee
\be
\frac{dM}{dt}=-b \tilde\alpha M
\label{eq:3}
\ee
\be
\frac{d Y}{dt}=\sum_i2C_2(R_i) Y \tilde{\alpha}-N^Y Y^2
\l{eq:4}
\ee
\be
\frac{dm_i^2}{dt}=4C_2(R_i) \tilde\alpha
M^2-YN^{m}_i(\sum_jm_j^2+A^2)
\l{eq:5}
\ee
\be
\frac{dA}{dt}= \tilde\alpha \sum_i2C_2(R_i) M-YAN^Y
\l{eq:6}
\ee
where $t=ln(\mu_0^2/\mu^2)$, $Y=\frac{h^2}{4\pi^2},\;
\tilde\alpha=\frac{g^2}{4\pi^2}$, $C_2(R_i)$ is the Casimir
appropriate
to the representation $R_i$ ($C_2(R)=\frac{N^2-1}{2N}$ for R the
fundamental representation of $SU(N)$) and $N^Y,\;N^m_i  $ counts the
number of independent (wave function renormalisation) diagrams
associated with the particular term.

To exhibit the fixed point structure it is convenient to form the
ratios $Y/\tilde\alpha$, $m/M$ and $A/M$. Then we easily find
\be
\frac{dln(Y/\tilde\alpha)}{d t}=(\sum_i
2C_2(R_i)+b)\tilde\alpha-N^Y_I
Y
\label{eq:yrg}
\ee
\be
\frac{d(m_i^2/M^2)}{d t}=4C_2(R_i)\tilde\alpha+2b\tilde\alpha
m_i^2/M^2-YN _i^m \frac{X}{M^2}
\l{eq:8}
\ee
\be
\frac{d (A/M)}{dt}= \tilde\alpha \sum_i2C_2(R_i)
+(2b\tilde\alpha-YN^Y
)(A/M)
\l{eq:9}
\ee
where $X=(\sum_jm_j^2+A^2)$.
One may immediately see that if $(\sum_i2C_2(R_i)+b)$ is positive
there
is an infra-red stable fixed point (IRSFP)
in the ratio $(Y/\tilde\alpha)$ given by

\be
\left(\frac{Y}{\tilde\alpha}\right)^*=\frac{(\sum_i2C_2(R_i)+b)}{N^Y
}
\label{eq:irfp1}
\ee
where we adopt the convention that fixed points are denoted with a *
superscript. In \cite{lr} we investigated the rate of approach to
such
a fixed point and concluded that it could be so rapid in many
extensions of the \sm that the ratio of couplings would be driven
very
close to the fixed point even though the range over which the
extension
of the \sm applied is very small (between $M_{Planck}$ and
$M_X\approx
10^{16}GeV)$.
Here we note that the ratio of soft SUSY breaking mass terms may
similarly be driven to IRSFP.
The ratio $(A/M)$ has an IRSFP given by

\be
\left(\frac{A}{M}\right)^*=1
\l{eq:12}
\ee
To determine the fixed point structure of the masses we assume that
eqs(\r{eq:irfp1},\r{eq:12}) apply and substitute them in
eqs(\r{eq:8},\r{eq:9}) to rewrite the RG equations in the form
\bea
\frac{d(X/M^2)}{dt} = 2\tilde\alpha
(2\sum_iC_2(R_i)-b)-((2\sum_iC_2(R_i)+b)\frac{\sum
N_i^m}{N^Y}-2b)(X/M^2)\tilde\alpha\nn
\frac{d(N_j^mm_i^2-N_i^mm_j^2)/M^2}{dt} =
2\tilde\alpha(2(N_j^mC_2(R_i)-N_i^mC_2(R_j)
+b(N_j^mm_i^2-N_i^mm_j^2)/M^2)
\l{eq:difficult}
\eea
 If $((2\sum_iC_2(R_i)+b)\frac{\sum N_i^m}{N_I^Y}-2b)$ is positive
the
ratio $(X/M^2)$ has an IRSFP given by
\be
\left(\frac{X}{M^2}\right)^*=\frac{4N^Y \sum
_iC_2(R_i)}{\sum_iN^m_i\sum_i2C_2(R_i)+(\sum_iN^m_i-2bN^Y)b}
\label{eq:m2}
\ee
Turning to the second of eqs(\r{eq:difficult}) we see it has a fixed
point if $b<0$ at
\be
\frac{N_j^mm_i^2-N_i^mm_j^2}{M^2}=
\frac{2(N_j^mC_2(R_i)-N_i^mC_2(R_j)}{(
-b)}
\ee

A further point of interest is the rapidity with which the couplings
and masses are driven to their fixed points. In \cite{lr} we gave an
analytic formula for the rate of approach of the Yukawa couplings and
concluded that in many interesting models it was very rapid; hence
our
interest in fixed points. For the masses the analytic solution of the
renormalisation group equations \cite{il} allows us also to compute
the
rate of approach. The analytic result is very lengthy and will be
published elsewhere; we will give the results in the specefic cases
discussed later.

\bigskip
\noindent{\bf Flavour changing processes.}
\bigskip

As we shall see, in many models of interest the conditions for these
IRSFP
are met. In them the number of independent parameters at the fixed
point is
dramatically reduced;in the model presented above there are just two
parameters, the gauge
coupling and the gaugino mass. This offers the possibility of
predicting quark
and lepton masses and mixing angles simply from the form of the
couplings
allowed by the symmetries of the theory. In order to generate mass
structures it is likely that some {\it
interactions} must be included which distinguish between generations
and
this raises a major difficulty in supersymmetric theories, namely the
difficulty
of controlling the flavour changing processes driven by the
non-degeneracy
of squark masses and the flavour changing $A_j$ terms of eq(\r{eq:1})
which arise as result of radiative corrections from these new
interactions.  The experimental limits on flavour changing processes
puts stringent limits on the squark and slepton masses of different
generations and on the $A_j$ terms\cite{hkr,hkr1}. The limits depend
on
the relative values of the  soft supersymmetry breaking terms in a
complicated way that is clearly discussed in \cite{hkr1}. For example
in the case that the ratio of gaugino masses to scalar masses is
small
at the Planck scale the limits restrict the off diagonal  squark and
slepton masses at $M_{GUT}$ (in the squark ``current" basis) of the
first two generations to be much less than their mean values

\begin{equation}
\frac{m_{\tilde{d}}^2-m_{\tilde{s}}^2}{m^{2(d)}_{av}}
\le  8.10^{-3}\left(\frac{M^{(d)}_{av}}{1TeV}\right)^2, \;\;\;\;
\frac{m_{\tilde{e}}^2-
m_{\tilde{\mu}}^2}{m^{2(l)}_{av}}
\le  10^{-1}\left(\frac{M_{av}^{(l)}}{1TeV}\right)^2
\label{eq:fc}
\ee
where $m_{av},\;M_{av}$ refers to an average value taken at the low
and
high (GUT) scales respectively. Since these off diagonal terms are
related to squark and slepton mass squared differences (in the squark
mass basis) times mixing angles we see the requirement for some
degree
of universality in the scalar masses. Since our proposal is that
IRSFP
dominate through
large radiative corrections we cannot rely on initial conditions to
ensure this degeneracy and to avoid flavour changing neutral currents
and thus it important to determine the
expectation for squark mass degeneracy at the IRSFP for models of
phenomenological interest.

The model presented above is very simple but it already sheds some
light on
this question. In particular eq(\r {eq:m2}) shows that the soft
squark
masses
are driven to fixed points determined simply by their representation
under the
gauge group. Thus family independent gauge interactions actually
drive the squark and slepton masses of different families to be equal
at the fixed point
irrespective of their initial values, reducing the generation of
favour
changing
neutral currents at low emerges. One explanation for the observed
smallness
of the latter is therefore that a family independent gauge group
dominates at
high energies.
However models capable of generating reasonable
fermion masses and mixings probably need to extend the gauge group to
include a family symmetry which, when broken, distinguishes between
different generations. The above discussion suggests a general
condition for
this to be true to the accuracy require: We consider a gauge group
which
has a product of factors some commuting with family indices and some
family
dependent. If there is some unification of these gauge factors either
in the
Grand Unified sense or in the superstring sense where there need not
be
a
single gauge group one may suppose that the gauge couplings are
comparable at the initial scale (the Planck scale or some Grand
Unified
scale
above the scale at which the couplings unify). However at low scales
these
couplings may diverge due to different radiative corrections. If
these
corrections make the gauge couplings associated with the family
symmetry
smaller than the family independent couplings then the latter will
dominate
the RG evolution of the soft masses and the squark masses may indeed
be
driven towards degeneracy. A particularly interesting example of this
is if the
family gauge symmetry is  Abelian. In this case the group is not
asymptotically
free so the coupling is driven {\it smaller} at low scales. Moreover
the beta
function for such a group is typically very large as the multiplicity
of terms can
be very large (for example left-handed squarks charged under the
group
contribute a multiplicity of 6 corresponding to three colours and two
flavours)
and so the  Abelian family symmetry gauge coupling can be very much
smaller than the family independent non- Abelian gauge coupling. We
shall discuss explicit realisations of this possibility in a
subsequent
paper.

Having discussed the possibility that there are family dependent
gauge
couplings we turn now to the question that there are family dependent
Yukawa couplings. To do this we must extend the model discussed above
to
include more than one family of quarks. Thus we consider the general
Yukawa couplings of the form $h_{ijk}Q_{i,L}Q^c_{j,R} H_k$ where we
have
assumed $n_g\times n_g$  independent Higgs fields coupling in
different
terms of the mass matrix. Now the RG equations for the Yukawa
couplings
are simple
generalisation of eq(\r{eq:yrg})
\be
\frac{d ln(Y_{ijk}/\tilde\alpha)}{dt}=(\sum_i2C_2(R_i)+b)
\tilde{\alpha}-
\sum_{r=i,j,k}n_r\sum_{s,t}\sqrt{Y_{rst}Y^*_{rst}})
\label{eq:ijk}
\ee
Again for $\sum_i(2C_2(R_i)+b) \tilde{\alpha}>0$ this has an IRSFP
for

\be
\left(\frac{Y_{ijk}}{\tilde\alpha}\right)^*=
\frac{\sum_i2C_2(R_i)+b}{N_T}
\ee
where $N_T=(n_H+n_{Q_L}+n_{Q_R})$ and $n_i$ counts the total number
of
wave function renormalisation graphs associated with the state i for
a
particular term in the RG equation. The point to
note is that, since the fixed point is being driven by a family
independent
gauge symmetry, the Yukawa couplings are similarly driven to a family
independent IRSFP. As a result, independent of the initial
conditions,
the
theory suppresses family dependence in the Yukawa couplings too. In
turn the soft SUSY breaking terms will also be driven to family
independent fixed points. The A terms have the form
$A_{ijk}Q_{i,L}Q^c_{j,R} H_k$. Since only wave function
renormalisation
enters the generalisation of eq(\r{eq:6}) is straightforward and has
a
fixed point $(A_{ijk}/M)^*=1$. Of course in a viable model all but
one
of the Higgs fields $H_k$ must acquire mass at some large scale $M$.
The  remaining field to be identified with the light MSSM Higgs field
will be some combination of $H_k$  and its couplings to the quarks
will
generate the quark mass matrix after electroweak breaking. The
essential feature of the IRSFP structure just discussed is that the
rotation that diagonalises the quark mass matrix will also
diagonalise
the A terms since both the Yukawa couplings $h_{ijk}$ and the A terms
$A_{ijk}$ have the same (family symmetric) form.  Similarly the soft
squark masses are driven to the degenerate family independent IRSFP.
Thus at the IRSFP the flavour changing processes induced by the soft
terms vanish. Of course if the gauge couplings are family
dependent this conclusion is not true. However if, as discussed
above,
the
family dependent coupling is driven to be negligible in the infra-red
then its effects on the Yukawa couplings will also be negligible
close
to the fixed point leaving the family independence unchanged.

It is perhaps worth emphasising two features of this example that
ensure the suppression of flavour changing processes. Firstly it is
important  that the structure of Yukawa couplings allows for a family
independent fixed point; if there had only been a single Higgs with
large coupling to the top quarks only, then only the top
squark would have had mass determined by a fixed point; the other
squark's mass
would have been radiatively corrected by the gauge couplings but
would
be
determined by their initial values and would not be driven to a
family
independent point. The same applies to the soft A terms.  In the
Yukawa
sector the important point is that the initial form of the Yukawa
couplings has, for special values of the Yukawa couplings, an
enhanced
family symmetry. This symmetry is then realised at the IRSFP driven
there by the radiative corrections of the family independent gauge
interactions. In the case discussed here we see there is indeed a
rich
$SU(3)_L\otimes SU(3)_R$ family symmetry at the fixed point under
which
$Q_L\;Q_R$ and $H$ transform as $(\bar{3},1)$ $(1,3)$ and
$(3,\bar{3})$
respectively. Of course it is not necessary for the symmetry to be so
large; it is easy to construct examples with just a permutation
symmetry acting on the quarks.
Secondly, the D-term associated with the (Abelian) family symmetry is
quite dangerous for flavour changing suppression  for it generates
squark and slepton masses and since the gauge factor is family
dependent these masses will be family dependent too. In order for the
family gauge symmetry to be broken at a high scale it is necessary
for
the scalar fields, $\phi,\;\bar{\phi}$, breaking this symmetry to
acquire vevs along a D-flat direction, $<\phi>=<\bar{\phi}>$.
Consider
first the usual radiative breaking mechanism under which radiative
corrections drive their soft masses squared,  $m^2,\; \bar{m^2}$
negative at some scale of $O(V)$ through radiative corrections. Then
it
is easy to show that $<\phi>\approx <\bar{\phi}>=O(V)$ and the D-term
vev is $g^2<\phi^2-\bar{\phi}^2>\propto (m^2-\bar{m}^2)$.  The masses
given to the squarks and sleptons from the D-term are thus
proportional
to $(m^2-\bar{m}^2)$ evaluated at the scale $V$ which, if there is no
cancellation or suppression, is of the order of the supersymmetry
breaking scale and potentially too large. In the case a single scalar
vev is induced by a Fayet Iliopoulos term \cite{fi} the situation is
even worse for the D-term is proportional to $m^2/g^2$ with no
possibility of cancellation. However the IRFP structure is such that
there are good reasons why the D-tem masses may not be too large. For
the first case it is quite likely that the pattern observed above
will
be repeated for the Yukawa couplings involving the $\phi$ and
$\bar{\phi}$ fields and that they will be driven\footnote{Provided
that
a symmetric structure of couplings involving the $\phi$ and
$\bar{\phi}$ fields exists} by the gauge coupling terms in the RG
equations (which are equal for conjugate representations) to equal
values at an IRSFP. In this case we can even allow for {\it
different}
initial values of $m$ and $\bar{m}$ for they too will be driven by
the
gauge couplings to be equal at an IRSFP. Even in the case there is
only
one scalar the flavour changing problem need not be severe because
the
$\phi$ (and $\bar \phi$) masses may be driven much smaller than the
squark and slepton masses. This will happen for example in the case
of
the Abelian family symmetry discussed above because its gauge
coupling
and gaugino mass are driven very small relative to the \sm couplings
and gaugino masses due to its large beta function. As may be seen
from
the discussion presented below the gaugino sets the scale for the
scalar mass at the IRSFP and hence the \sm singlet fields will have
very small scalar masses as required.

To summarise, we have demonstrated how IRSFP determine both Yukawa
couplings and soft SUSY breaking terms and shown there exist a class
of
models which, even with the introduction of family dependent gauge
couplings, may still lead to an effective low-energy lagrangian which
has
family independent couplings and soft SUSY breaking masses. This
opens
the way to construct models in which the pattern of fermion masses
and
mixing angles is determined by an underlying gauge symmetry yet which
does not lead to large flavour violation at low energies induced by
squark and
slepton masses \cite{gr}. Moreover in such theories there may only be
two
independent parameters, the dominant gauge coupling and one gaugino
mass allowing for the possible prediction of both the structure and
{\it
magnitude} of fermion masses and mixings.

We turn now to a discussion of these ideas in the context of specific
models. However we will confine our attention in this paper to the
expectation for the relative sizes of SUSY breaking terms following
from the IRSFP structure\footnote{This will allow us to address an
outstanding question, namely the implication of the infra red
structure of the theory
for the $\mu^2$ term which couples the superfields $H_1$ and
$H_2$ in the superpotential of the MSSM where $H_{1,2}$ contain the
Higgs
scalars giving the down and up quark masses respectively.} .
 A study in detail of the flavour violation to be expected requires
the
construction of a viable theory of fermion masses and this we will
discuss in another paper.

\bigskip
\noindent{\bf IRSFP in the MSSM}
\bigskip

In \cite{lr} we considered the IRSFP structure for the top Yukawa
coupling in the MSSM. In the approximation of ignoring the
$SU(2)\otimes U(1)$ couplings the theory the RG equations for the
soft
SUSY breaking terms are \cite{st1,st2}
\bea
\frac{dM}{dt} &=& -b_3\ta3 M \nn
\frac{dB}{dt} &=& - 3 a \ta3 A_t \nn
\frac{d\tilde A_t}{dt}&=& \ta3 (\fr -(6a-b_3)\tilde A_t)\nn
\frac{d(\tilde\mu_2^2-\tilde\mu^2)}{dt}&=&\ta3(-3a\tilde X+
2b_3(\tilde\mu_2^2-\tilde\mu^2)) \nn
\frac{d \tilde m_{t_r}^2}{dt}&=& \ta3(\fr-2a\tilde X+2b_3\tilde
m_{t_r}^2) \nn
\frac{d\tilde m_{Q_t}^2}{dt}&=&\ta3(\fr-a\tilde X+2b_3\tilde
m_{Q_t}^2)
\nn
\frac{d\tilde\mu_3^2}{dt} &=& \ta3(3\tilde Aa+b_3\tilde\mu_3^2)\nn
\frac{d\mu^2}{dt} &=& -3Y_t\mu^2
\l {eq:sum}
\eea
where $a=(\frac{Y_t}{\ta3})^*$ and the $\tilde{}$ means the masses
normalised to $M^2$, the running gluino mass, except in the
case of $\mu_3^2$ which is normalised to $M_3\mu$. Also

\be
\tilde X=(m_{Q_t}^2+m_{t_R}^2+\mu_2^2-\mu^2+A_t^2)/M^2
\ee

As discussed above to solve these equations for their IRSFP structure
it is best to take linear combinations and rewrite the RG equations
in
the form
\bea
\frac{d\tilde X}{dt}&=&\ta3(2\fr-6a\tilde X+2b_3 \tilde X+2\fr
\tilde A_t-12a\tilde A_t^2)\nn
\frac{d (\tilde m_{t_r}^2-2\tilde m_{Q_t}^2)}{dt}&=&
\ta3(-\fr+2b_3(\tilde m_{t_R}^2-2\tilde m_{Q_t}^2) \nn
\frac{d(\tilde\mu_2^2-\tilde\mu^2-3\tilde
m_{Q_t}^2)}{dt}&=&\ta3(-3\fr+
2b_3(\tilde\mu_2^2-\tilde\mu^2-3\tilde m_{Q_t}^2))
\l{eq:rgp}
\eea
 In the case of the MSSM $b_3 = -3$ and in the neighbourhood if the
IRSFP $a\approx a^*=\frac{7}{18}$ so we see that the condition for
infra-red stability of {\it all} the soft masses is satisfied.
Solving
for the positions of these fixed points gives
\bea
\tilde A_t^{*} &=& 1 \nn
\tilde X^* &=& 2 \nn
(\tilde\mu_2^2-\tilde\mu^2)^* &=& -\frac{7}{18} \nn
\tilde m_{t_R}^{2*} &\approx&\frac{34}{54} \nn
\tilde m_{q_t}^{2*} &\approx&\frac{41}{54}\nn
\tilde \mu_3^{2*} &=& \frac{7}{18}\nn
\mu^2&=&0
\l{eq:mssm}
\eea
Will these fixed points be relevant to the determination of the
physical
parameters?  To answer this we have determined the low energy
parameters
 using the complete analytic solution of the renormalisation group
equations \cite{il}. As discussed in \cite{lr} in the limit that at
 the gauge coupling unification scale the ratio of the top Yukawa
coupling to the gauge coupling is initially much larger than the
 fixed point value there is a calculable radiative correction to
the fixed point value at low energy. This gives Hill's ``quasi" fixed
point value \cite{hill} roughly a factor of 2 larger than the true
fixed
 point value. For the soft SUSY breaking parameters we have
calculated
the
 low energy values assuming the initial values of the top Yukawa
coupling and soft terms are in the domain of attraction of the fixed
point.

The rate of approach to the fixed points can be seen from the
following
approximate expressionsformed by expanding the analytical solution
around the
fixed point values.

\bea
\at &=& 0.98 + 0.14 ( \aa - 1 ) - (209.55 - 70.33(\aa-1))\y \nn
\xt &=& 1.85 + 0.07 (\aa-1) + 0.01(\mh+0.39) + 0.10(\mq-0.76) \nn
    & &+0.05(\mt-0.63)+(-529.68-149.89(\aa-1)-39.96(\mh+0.39) \nn
    & &-13.32(\mq-0.76) -26.64(\mt-0.63))\y\nn
\frac{B(t)}{M(t)} &=& -0.29 - 0.12(\aa-1) + 0.38 (\frac{B}{M} + 0.38
) \nn
& &- ( 104.77 +  35.16(\aa-1))\y\nn
\mqt &=& 0.75 -0.03(\aa-1)+0.10(\mq-0.76)\nn
     & &+(-27.90+7.52(\aa-1) - 13.32(\mq-0.76))\y\nn
\mtt &=& 0.51 - 0.07(\aa-1) +0.05(\mt-0.63)\nn
     & & +(-52.34+15.03(\aa-1)-26.64(\mt-0.63))\y\nn
\frac{(\mu_2^2-\mu^2)(t)}{M(t)^2} &=& -0.39 - 0.10(\aa-1)
+0.01(\mh+0.39)\nn
        & & +(-37.81+22.55(\aa-1)-39.96(\mh+0.39))\y\nn
\eea

 From the first equation we see that a 100 \% deviation in the
initial
value of A from its fixed point value results in only a 14\%
deviation
at low energies. However the sensitivity of A to the initial value of
the Yukawa coupling is greater. Using an initial value of $\tilde
\alpha =1/(25 \pi)$ a 100\% increase of $Y_0$ from its fixed point
corresponds to a change of -1 in (A/M). This may be expected to be
the
minimum possible change if the Yukawa coupling lies at its quasi
fixed
point. However as the remaining equations show the sensitivity to
deviations of the initial values from the fixed point values are
smaller for the other quantities. This completes our discussion of
the
MSSM fixed point structure, some aspects of which has been discussed
by
other authors\cite{st2}. However as discussed above our expectation
is
that the IRSFP structure will be even more relevent for models beyond
the \sm. To illustrate the expectations we consider the three models
discussed in \cite{lr} based on the unification groups $SU(3)^3$,
$SU(5)$ and $SO(10)$

\bigskip
\noindent ${\bf SU(3)^3}$
\bigskip

The group $SU(3)^3$ has many attractions as a non-Grand-Unified
extension of the \smp Provided the multiplet content is chosen
symmetrically, the gauge couplings above the $SU(3)^3$ unification
scale, $M_X$, evolve together. As a result it offers an example of a
theory which can be embedded in the superstring which preserves the
success of the unification predictions for gauge couplings even if,
as
has been found to be usually the case in the compactified string
theories so
far analysed, the unification or compactification
scale is much higher than $10^{16}GeV$.  The light multiplet
structure
after
symmetry breaking we take to be just that of the MSSM.  The $n_g$
families are contained in $n_g$ copies of $I$
representations where
$I=((1,3,\bar{3})+(\bar{3},1,3)+(3,\bar{3},1))$.
In addition there are two further copies of $I$ representations which
contain the Higgs fields.  Taking a single dominant Yukawa coupling
its
fixed point is given by Table \r{table:1} for n=0
The fixed point for the A
term may be immediately determined giving

\be
\left(\frac{A }{M}\right)^*=1
\l{eq:su33}
\ee

Up to now we have omitted any discussion of the $\mu$ term in this
model.  these terms. The $SU(2)$ doublet supermultiplets containing
the
Higgs fields
$H_{1,2}$ giving rise to down and up quarks reside in the
$(1,3,\bar{3})$
representation
\be
(1,3,\bar{3})=\left(\begin{array}{ccc}
H_1 H_2 L\\
E^C \nu_r N
\end{array}\right)
\ee
where $L$ is a lepton doublet supermultiplet, $E^C$ is the charge
conjugate
singlet charged lepton supermultiplet and $\nu_R $ and $N$ are
supermultiplets
neutral under the \smp The $\mu$ term comes from the coupling
$\lambda\epsilon_{abc}\epsilon^{ijk}A^a_i B^b_j C^c_k$ where $A,\;B$
and
$C$ are superfields (assumed distinct here) containing $H_1,\;H_2$
and
the
$N$ scalar which acquires a vev generating the $\mu$ term,
$\mu=\lambda<N>$. There are two extremes for the magnitude of this
vev.
It
could be of $O(M_W)$ if the N field remains light. This model
corresponds to
the (M+1)SSM which in our opinion is disfavoured because of
cosmological
problems due to the domain wall production associated with the
discrete
symmetry present in such models\cite{able}. Alternatively it could be
of $O(M_X)$ (the
maximum it could be since it breaks $SU(3)^3$). In this case
$\lambda$
must
be very small, of $O(M_W/M_X)$. There is a very plausible origin for
this\cite{gm} for
if the term proportional to $\epsilon_{abc}\epsilon^{ijk}A^a_i B^b_j
C^c_k/M_{Planck}$ is present in the Kahler potential. When
supersymmetry is
broken a term proportional to
$m_{3/2}\epsilon_{abc}\epsilon^{ijk}A^a_i
B^b_j C^c_k/M_{Planck}$ is generated in the superpotential
corresponding to $\lambda=O(m_{3/2}M_X/M_{Planck}$). We will assume
this
is the mechanism that is operative in which case, due to the
smallness
of
$\lambda$, the RG equations are little changed by loops generated by
$\lambda$ couplings. The things that do change
are due to the induced $\mu H_1 H_2$ term. Although the N field does
not
acquire a VEV until the scale $M_X$, radiative corrections generating
terms with external N fields
involving loop momentum above this scale are important for the
external
fields do not ``feel" the loop momenta and may be replaced by their
vevs.

Now we can discuss the RG equations for the squark mass and
$\mu,\;m_1^2, \;m_2^2$ and $m^2_3$ where the last three masses are
the
coefficients of the $\mid H_1\mid^2$, $\mid H_2\mid^2$ and $H_1H_2$
terms
in the scalar potential. For the purpose of illustration it is
sufficient to return to
our one generation example and assume that only the top Yukawa
coupling
is
significant. The RG equations are
\bea
\frac{dM}{dt}&=&-6\tilde\alpha M \nn
\frac{dm_{\tilde q}^2}{dt}&=&\frac{32}{3}\tilde\alpha M^2-
3Y(2m_q^2+\mu_2^2+A^2-\mu^2)\nn
\frac{d\mu^2}{dt}& =&(\frac{32}{3}\tilde\alpha-3Y)\mu^2\nn
\frac{d\mu_2^2}{dt}&=&\frac{32}{3}\tilde\alpha(M^2+\mu^2)-
3Y(\mu_2^2+2m_q^2+A^2)\nn
\frac{d\mu_1^2}{dt}&=&\frac{32}{3}\tilde\alpha(M^2+\mu^2)-3Y\mu^2\nn
\frac{d\mu_3^2}{dt}&=&(\frac{16}{3}\tilde\alpha-\frac{3}{2}Y)\mu_3^2-
\frac{32}{3}M\mu\tilde\alpha+3\mu A Y
\l {eq:su3}
\eea

This has IRSFP
\bea
\left(\frac{M}{\mu}\right)^*=0\nn
\left(\frac{2m_{\tilde q}^2+\mu_2^2-\mu^2}{\mu^2}\right)^*=0\nn
\left(\frac{\mu_{1,2}^2}{\mu^2}\right)^*=1\nn
\left(\frac{\mu_3^2}{\mu^2}\right)^*=0,\;\;
\left(\frac{\mu M}{\mu_3^2}\right)^*=0
\eea

Thus an hierarchy of masses develops with $\mu,\;\mu_{1,2}$ and
$\mu_3$
growing roughly
 at the same rate while $m_{\tilde q}^2$ and $\mu_2^2-\mu^2$ tend to
constant values
 and $M$ and $A$ fall towards zero. In this case only $A$ is in the
domain of attraction
of the \sm fixed point (the $SU(3)^3$ fixed point is actually at it).
However the combination
of masses $X=2m_{\tilde q}^2+\mu_2^2-\mu^2$ {\it does} also lie in
the
domain of attraction
of the MSSM fixed point and should also closely approach it.

\begin{table}
\begin{center}
\begin{tabular}{|c|c|c|c|}\hline
$n$  &  $(\frac{Y_t}{\tilde\alpha})^*$ & \EPP  &
$(\frac{\tilde\alpha}{Y_t})_{SU(3)^3}^*/
(\frac{\tilde\alpha}{Y_t})_{MSSM
}^*$
\\ \hline\hline
0    & 2.44 & 0.48 & 0.16 \\ \hline
4    & 5.11 & 0.02 & 0.08 \\ \hline
\end{tabular}
\end{center}
\l{table:1}
\caption{$SU(3)^3$}
\end{table}

The rate of approach to this fixed points can be seen from the
following expressions taken from the analytical solution to the
renormalization group equations. Expanding the solution around the
fixed point values we get:

\bea
\at &=& 1 + 0.62 ( \aa- 1 ) - ( 40.13 + 32.09 (\aa - 1) ) \y  \nn
\xtmu &=& 0.47 \mtmu + 0.47 \mqmu + 0.47 \mhmu + 0.18 \ammu
           + 0.59 \mmu \nn
         & & - ( 16.17 \mtmu + 16.17 \mqmu + 16.17 \mhmu + 18.93
\ammu
          + 5.44 \mmu) \y \nn
\mqtmu &=& 0.47 \mqmu - 0.04 \ammu
           + 0.17 \mmu - ( 16.17 \mqmu + 0.78 \ammu - 1.05 \mmu ) \y
\nn
\mhtmu &=& 1 + 0.47 \mhmu - 0.04 \ammu
          + 0.17 \mmu - ( 16.17 \mqmu + 0.78 \ammu  - 1.05 \mmu) \y
\nn
\eea

The rate of approach to the fixed point is not very great. However if
we add to our theory n copies of chiral superfields in
$(I+\bar I)$ representations things change dramatically as was
discussed in \cite{lr}. The factor determining the closeness of
approach to the fixed point of the Yukawa couplings is given by the
third column of Table \ref{table:1} and for n=4 the approach is
within
2\% compared to 48\% for n=0. The reasons were discussed in \cite{lr}
but largely follow from the fact that the additional matter makes the
couplings run fast.  For the case of $n=4$ the rate of approach to
the
fixed point is given by the following expressions:

\bea
\at &=& 1 + 0.39 ( \aa- 1 ) - ( 5.40 + 3.97 (\aa - 1) ) \y  \nn
\xtmu &=& 0.09 \mtmu + 0.09 \mqmu + 0.09 \mhmu + 0.04 \ammu
           + 0.11 \mmu  \nn
       & &   - ( 0.61 \mtmu + 0.61 \mqmu + 0.61 \mhmu - 0.18 \ammu
           + 0.03 \mmu) \y \nn
\mqtmu &=& 0.09 \mqmu - 0.02 \ammu
            +0.03 \mmu - ( 0.61 \mqmu - 0.12 \ammu  + 0.003 \mmu) \y
\nn
\mhtmu &=& 1 + 0.09 \mhmu - 0.02 \ammu +0.03\mmu
- ( 0.61 \mqmu  - 0.12 \ammu + 0.003 \mmu ) \y \nn
\eea

It can be seen that the effect of additional matter dramatically
speeds
up the rate of approach to the fixed points of the soft terms too.

\bigskip
\noindent ${\bf SU(5)}$
\bigskip

\begin{table}
\begin{center}
\begin{tabular}{|c|c|c|c|} \hline
$n$ & $(\frac{Y_t}{\tilde\alpha})^*$ & \EPP &
$(\frac{\tilde\alpha}{Y_t})_{SU(5)}^*/
(\frac{\tilde\alpha}{Y_t})_{MSSM}^
*$
\\ \hline\hline
0 & 1.80 & 0.62 & 0.22 \\ \hline
4 & 3.57 & 0.29 & 0.11 \\ \hline
\end{tabular}
\end{center}
\l{table:2}
\caption{SU(5)}
\end{table}

In our next example we will consider an SU(5) model in which matter
is arranged in three generations in $I$=
$\{ \psi^{xy}(10) + \phi_x(\overline{5})\}$ representations together
with n further copies of chiral superfields in ($I+\bar{I}$)
representations plus a Higgs sector made
up of a (complex) adjoint, $\Sigma(24)$, to break SU(5) and a set of
Weinberg-Salam 5-plets $H_1(5) + H_2(\overline{5})$. We consider the
case of a single large Yukawa coupling leading to the top quark mass.
The Yukawa coupling IRFP structure and the rate of approach is
summarised in Table \ref{table:2} \cite{lr}. For brevity we only give
here  the expansion about the fixed point for the soft terms for the
case n=4 :
\bea
\at &=& 1 + 0.37 ( \aa- 1 ) - ( 33.66 + 21.82 (\aa - 1) ) \y  \nn
\xtmu &=& 0.19 \mtmu + 0.19 \mqmu + 0.31 \mhmu + 0.04 \ammu + 0.56
\mmu
\nn
& & - ( 11.24 \mtmu + 11.24 \mqmu + 6.24 \mhmu + 9.69 \ammu +
2.86\mmu)
\y \nn
\mqtmu &=& 0.18 \mqmu - 0.06 \ammu + 0.19 \mmu
 - ( 11.24 \mqmu - 1.23 \ammu + 0.01 \mmu) \y \nn
\mhtmu &=& 1 + 0.31 \mhmu - 0.05 \ammu + 0.08 \mmu
- ( 6.24 \mqmu  - 0.98 \ammu + 1.42 \mmu) \y \nn
\eea

As in the $SU(3)^3$ case the effect of additional matter
is to speed up the approach to the fixed point for the soft terms
too..

\bigskip
\noindent ${\bf SO(10)}$
\bigskip

\begin{table}\begin{center}
\begin{tabular}{|c|c|c|c|} \hline
$n$ & $(\frac{Y_t}{\tilde\alpha})^*$ & \EPP &
$(\frac{\tilde\alpha}{Y_t})_{SO(10)}^*/
(\frac{\tilde\alpha}{Y_t})_{MSSM}
^*$
\\ \hline\hline
0 & 1.75 & 0.51 & 0.22 \\ \hline
4 & 2.89 & 0.24 & 0.13 \\ \hline
\end{tabular}
\end{center}
\l{table:3}
\caption{SO(10)}
\end{table}

Finally for $SO(10)$ with three families and n=4 vectorlike
($16\;+\;\bar{16}$) representation the IRSFP structure and the rate
of
approach to it is summarised in Table \ref{table:3} and the following
equations. In this example we have included the additional $16$, $45$
and $10$ dimensional vectorlike representations needed to break the
group.

\bea
\at &=& 1 + 0.37 ( \aa- 1 ) - ( 33.66 + 21.82 (\aa - 1) ) \y  \nn
\xtmu &=& 0.19 \mtmu + 0.19 \mqmu + 0.31 \mhmu + 0.04 \ammu + 0.76
\mmu
\nn
& & - ( 11.24 \mtmu + 11.24 \mqmu + 6.24 \mhmu + 9.69 \ammu +
6.29\mmu
) \y \nn
\mqtmu &=& 0.18 \mqmu - 0.06 \ammu + 0.21 \mmu
- ( 11.24 \mqmu - 1.23 \ammu + 0.89 \mmu) \y \nn
\mhtmu &=& 1 + 0.31 \mhmu - 0.05 \ammu + 0.16 \mmu
- ( 6.24 \mqmu  - 0.98 \ammu + 0.71 \mmu) \y \nn
\eea
Again one may see the relatively rapid approach to the fixed points.

To summarise we have considered the implications for the soft SUSY
breaking terms of the IRSFP structure in a variety of models. We have
argued that such structure is likely to be relevant for a wide
variety
of models beyond the \sm with couplings and masses driven very close
to
the fixed point. A study of the family dependent effects shows that
in
a sub-class of models the resultant flavour changing neutral
processes
are very small at the fixed points even in theories with family
dependent interactions offering the possibility of constructing
viable
models of family interactions capable of generating the observed
fermion mass structure. We have determined the fixed point
predictions
for the soft SUSY breaking terms for a variety of models and also the
rate of approach to the fixed point structure. This shows that some
subset of the soft terms will be determined quite well by the fixed
point structure substantially limiting the parameter space of the
low-energy effective supersymmetric theory.

\noindent{\bf Acknowledgement.} One of us (GGR) would like to thank
M.Carena, S.Dimopoulos, L.Ibanez, H.Murayama and C.Wagner for useful
discussions.


\begin{thebibliography}{99}
\bibitem{pr}B.  Pendleton, G.  G.  Ross, Phys.  Lett.  98B,
291(\-1981);
\bibitem{lr} M. Lanzagorta and G. G. Ross, Phys. Lett. B349:319-328,
1995.
\bibitem{fn}
J. Harvey, P. Ramond and D. Reiss,
Phys.Lett.B92(1980)309; \\
M.E. Machacek and M.T. Vaughn, Phys. Lett.
{\bf B103} (1981) 427;\\
C. Wetterich, Nucl. Phys. {\bf B261} (1985) 461;
Nucl. Phys. {\bf B279} (1987) 711;\\
J. Bijnens and C. Wetterich,
Phys. Lett. {\bf B176} (1986) 431;
Nucl. Phys. {\bf B283} (1987) 237;
Phys. Lett. {\bf B199} (1987) 525;\\
 P. Kaus and S. Meshkov, Mod. Phys. Lett. {\bf A3}
(1988) 1251;\\
C.D.Froggat and H.B. Nielsen, Origin of symmetries, World
Scientific (1991);\\
J.L.Lopez and D.V.Nanopoulos, Phys. Lett. B268(1991) 359;\\
S. Dimopoulos, L. J. Hall and S. Raby,
Phys.Rev.Lett.68  (1992) 1984;
Phys.Rev. D45 (1992) 4195;\\ H. Arason,
D. J. Casta\~no, P. Ramond and E. J. Piard,
Phys.Rev.D47(1993)232; \\
G. F. Giudice, Mod. Phys. Lett. {\bf A7}
(1992)2429.
\bibitem{st2}M.Carena, M.Olechowski, S.Pokorski and C.E.M.Wagner,
Nucl.Phys.  B419 (1994) 213;  ibid. B426 (1994) 269.
\bibitem{j1}I.Jack and D.R.T.Jones,Phys,Lett.B349(1995)294;\\
I.Jack, D.R.T.Jones and K.L.Roberts, LTH 347 (hep-ph/9505242);\\
P.M.Ferreira, I.Jack and D.R.T.Jones, LTH 352 (hep-ph/9506467)
\bibitem{il}
L.E. Ib\'{a}\~{n}ez and C. L\'{o}pez, Phys. Lett. \underline{126B}
(1983) 54; Nucl. Phys.
\underline{B233} (1984) 511.
\bibitem{hkr} S. Bertolini and A. Masiero, Phys. Lett.
B174(1986)343;\\
R.Barbieri and L.J. Hall, Phys. Lett. B338(1994)212;\\
R.Barbieri, L.J.Hall and A.Strumia, preprint IFUP-YH 72/94 (1995).
\bibitem{hkr1}F.Gabbiani and A.Masiero, Nucl. Phys. B322(1989)235;\\
J.Hagelin, S.Kelley, T.Tanaka, Nucl.Phys.B415(1994)293;\\
Mod.Phys.Lett.A8(1993)2737;\\
D.Choudhury, F.Eberlein, A.Konig, J.Louis and S. Pokorski, Max-Planck
Institute preprint, MPI-PhT/94-51;
\bibitem{hill} C.  T.  Hill, Phys.  Rev.  D24, 691, (1981); C.  T.
H\-ill, C.  N.  Leung, S.  Rao, Nucl.  Phys.  B262, 517(1985);
\bibitem{fi}P.Fayet and J. Iliopoulos, Phys.Lett.51B(1974)461.
\bibitem{gr}G.G.Ross, CERN preprint CERN-TH 95-162 (1995)
\bibitem{st1}W.A.Bardeen, M.Carena, S.Pokorski and C.E.M.Wagner,
Phys.
Lett.B320(1994);\\
N.Krasnikov, G.Kreyerhoff and R.Rodenberg, Il Nuovo cimento,
108A(1995)565.\\
P.Langacker, N.Polonsky, Phys. Rev. D49(1994)1454;\\
G.K.Leontaris and N.D.Tracas, Ioannina Univ. preprint IOA 303/94,
NTUA
44/94\\
M.Carena and C.E.M.Wagner, CERN preprint CERN-TH.7320/94;
\bibitem{able}S.Abel, S.Sarkar and P.White, preprint hep-ph/9506359
\bibitem{gm}G.F.Giudice and A.Masiero, Phys. Lett. B206(1988)480.


\end{thebibliography}
\end{document}